# A guided stochastic energy-domain formulation of the second order Møller-Plesset perturbation theory

Qinghui Ge[†*], Yi Gao[†], Roi Baer[‡], Eran Rabani[♣] and Daniel Neuhauser[†]

[†] Department of Chemistry and Biochemistry, University of California, Los Angeles CA-90095, USA.
[*] Department of Chemistry, Zhejiang University, 38 Zheda Road, Hangzhou, China, 310027.
[‡] Fritz Haber Center for Molecular Dynamics, Institute of Chemistry, The Hebrew University of Jerusalem, Jerusalem 91904, Israel.
[♣] School of Chemistry, The Sackler Faculty of Exact Sciences, Tel Aviv University, Tel Aviv 69978, Israel.

**ABSTRACT:** We develop an alternative formulation in the energy-domain to calculate the second order Møller-Plesset (MP2) perturbation energies. The approach is based on repeatedly choosing four random energies using a non-separable guiding function, filtering four random orbitals at these energies, and averaging the resulting Coulomb matrix elements to obtain a statistical estimate of the MP2 correlation energy. In contrast to our time-domain formulation, the present approach is useful for both quantum chemistry and real-space/plane wave basis sets. The scaling of the MP2 calculation is roughly linear with system size, providing a useful tool to study dispersion energies in large systems. This is demonstrated on a structure of 64 fullerenes within the SZ basis as well as on silicon nanocrystals using real-space grids.

The second order Møller-Plesset (MP2) perturbation theory is one of the simplest and most fundamental forms used to introduce correlations in electronic structure calculations.[1] The formal MP2 expression can be manipulated into the following form:

$$E_{MP2} = -\sum_{ijab} \frac{\langle ab|ij\rangle^2 + \langle ab|ji\rangle^2 - \langle ab|ij\rangle\langle ij|ab\rangle}{\varepsilon_a + \varepsilon_b - \varepsilon_i - \varepsilon_j} \quad (1)$$

where $\langle ab|ij\rangle$ and $\langle ab|ji\rangle$ are the Coulomb and exchange matrix elements, respectively, $\varepsilon_s$ is the orbital energy, and the indices $i,j$ and $a,b$ refer to occupied and virtual states, respectively. Direct application of Eq. (1) involves CPU time scaling as $\mathcal{O}(N^5)$ with the size of system ($N$). Such scaling severely caps the size of systems which can be studied and serves as a driving force to develop more efficient computational schemes.[2-12] In these methods, linear scaling emerges once the system is larger than the one particle density matrix range, typically above a few tens of thousands of atoms, limiting the efficacy of the methods considerably.

In a previous paper [13] we developed an expeditious stochastic approach to obtain the MP2 energy, whereby the energy denominator in Eq. (1) is replaced by integration over a real-time correlation function and the exact eigenstates are replaced by arbitrary combinations of random states filtered to be in the occupied and virtual space, respectively. Application to hydrogen passivated silicon nanocrystals with thousands of electrons represented on a real-space grid provided accurate estimates of the MP2 energies in systems far beyond the capabilities of present day MP2 implementations. Other stochastic schemes have been recently proposed to reduce the computational workload of the MP2 calculation,[14,15] or other approaches to treatment of electron correlations.[13,16-18]

Here we present an alternative stochastic approach, akin to our previous work on multiexciton generation rates[19], more suitable for quantum chemistry basis-sets and nearly as efficient for grid-based representation. The approach is based on choosing four random energies using a non-separable guiding function and then filtering four random orbitals at these energies. The MP2 energy is then estimated as an average over different combinations of random orbitals, where for each combination we calculate the contribution to the MP2 energy given by Eq. (1) with a proper weight.

The motivation for the development of the present energy-domain method is to avoid the costly repeated estimates of the Coulomb matrix elements in the time domain formulation. In real-space/plane-wave basis this is relatively cheap, but when quantum-chemistry basis sets are used, the estimation of the Coulomb integral at each time step becomes the most demanding portion of the calculation, and thus, prohibits application to large systems. In contrast, the energy-domain formulation requires the evaluation of a single Coulomb integral for each set of random orbitals, providing a framework for a stochastic MP2 approach suitable for quantum chemistry basis sets.

To start, let us note the following relation, replacing a sum over states by an average $\langle ... \rangle_\chi$ over random orbitals $\chi_i(\mathbf{r})$:

$$\sum_{i\in occ} f(\varepsilon_i)\langle i|\hat{A}|i\rangle = \int f(\varepsilon) \left\langle \langle i_\varepsilon|\hat{A}|i_\varepsilon\rangle \right\rangle_\chi d\varepsilon \quad (2)$$

where:

$$i(\mathbf{r}) \equiv i_\varepsilon(\mathbf{r}) = \theta(\mu - \varepsilon) \sqrt{\frac{\delta_\sigma(\varepsilon - H)}{\rho(\varepsilon)}} \chi_i(\mathbf{r}) \quad (3)$$

is a "projected" random occupied orbital at energy $\varepsilon$, $\delta_\sigma(\varepsilon) = \frac{2}{\pi}\frac{\sigma^3}{(\varepsilon^2+\sigma^2)^2}$ is a squared Lorentzian filter function, $\rho(\varepsilon) = tr(\varepsilon - H)$ is the density of states (DOS), $\theta(x)$ is the Heaviside step function, and $\mu$ is the chemical potential. Similarly, for a virtual orbital we have an analogous expression with the following random orbitals:

$$a(r) \equiv a_\varepsilon(r) = \theta(\varepsilon - \mu)\sqrt{\frac{\delta_\sigma(\varepsilon - H)}{\rho(\varepsilon)}}\chi_a(r) \quad (4)$$

With this notation the MP2 energy can be written as:

$$E_{MP2} = -\int \frac{V^2(\{\varepsilon\})}{\varepsilon_a + \varepsilon_b - \varepsilon_i - \varepsilon_j} p(\{\varepsilon\}) d^4\{\varepsilon\}, \quad (5)$$

where $\{\varepsilon\}$ is shorthand notation for $\{\varepsilon_i, \varepsilon_j, \varepsilon_a, \varepsilon_b\}$ and $\{\chi\}$ for $\{\chi_i, \chi_j, \chi_a, \chi_b\}$, $p(\{\varepsilon\}) = \rho(\varepsilon_i)\rho(\varepsilon_j)\rho(\varepsilon_a)\rho(\varepsilon_b)$, and the positive definite coupling $V^2(\{\varepsilon\}) = \langle\langle ab|ij\rangle^2 + \langle ab|ji\rangle^2 - \langle ab|ij\rangle\langle ij|ab\rangle\rangle_{\{\chi\}}$. The integrand in Eq. (5) has contributions from the Coulomb matrix elements which depend on the $\varepsilon$'s and the density of states and the denominator which also depends on the $\varepsilon$'s in an obvious way. To perform the integral, we need to evaluate it using a Monte Carlo procedure with importance sampling based on a guiding function $w(\{\varepsilon\})$:

$$E_{MP2} = -\int \frac{V^2(\{\varepsilon\})p(\{\varepsilon\})}{(\varepsilon_a + \varepsilon_b - \varepsilon_i - \varepsilon_j)w(\{\varepsilon\})} w(\{\varepsilon\}) d^4\{\varepsilon\} \quad (6)$$

We find that a guiding weight of the form

$$w(\{\varepsilon\}) = p(\{\varepsilon\})(\varepsilon_a + \varepsilon_b - \varepsilon_i - \varepsilon_j)^{-\alpha}$$
$$\times \left(\frac{1}{\left((\varepsilon_a - \varepsilon_i) - (\varepsilon_b - \varepsilon_j)\right)^2 + \beta^2}\right.$$
$$\left. + \frac{1}{\left((\varepsilon_a - \varepsilon_j) - (\varepsilon_b - \varepsilon_i)\right)^2 + \beta^2}\right), \quad (7)$$

works quite well and the numerical fluctuations are fairly insensitive to the values of $\beta$ (chosen here as $\beta^2 = \frac{1}{100}$ and $\alpha = 3$). This function accounts for the fact that the coupling matrix elements tend to be large whenever the particle-hole pairs are close in energy, i.e., whenever $\varepsilon_b - \varepsilon_i \approx \varepsilon_a - \varepsilon_j$ or $\varepsilon_b - \varepsilon_j \approx \varepsilon_a - \varepsilon_i$.

We applied the energy-domain formalism using a Gaussian basis set, where the $M \times M$ overlap and Hamiltonian matrices, $S$ and $H$ ($M$ is the size of the basis) are transformed to an orthogonal basis, $\widetilde{H} = S^{-\frac{1}{2}}HS^{-\frac{1}{2}}$ and then we ize $\widetilde{H} = U\varepsilon U^T$. The density of states is constructed explicitly from the eigenvalues of $\widetilde{H}$, e.g.:

$$\rho(\varepsilon) = \sum_{s=1}^{M} \delta_\sigma(\varepsilon - \varepsilon_s). \quad (8)$$

Then, in each calculation we randomly sampled four energies from the weight function $w(\{\varepsilon\})$ and also chose four orbitals described by random coefficients (with respect to the basis set), $\chi_s$. These coefficients are uniformly generated in the range $-\sqrt{3} < \chi_s < \sqrt{3}$, where $s = 1, ..., M$. This choice ensures that the components of the random vectors are mutually orthonormal on average, i.e., $\langle\chi_s\chi_t\rangle_\chi = \delta_{st}$. The four random orbitals are then filtered, e.g. when the energy is smaller than $\mu$ one obtains an occupied random orbital:

$$(\bar{\iota}_\varepsilon)_s = \sqrt{\frac{\delta_\sigma(\varepsilon - \varepsilon_s)}{\rho(\varepsilon)}}\chi_s. \quad (9)$$

Similarly, for a virtual random orbital one obtains:

$$(\bar{\iota}_\varepsilon)_a = \sqrt{\frac{\delta_\sigma(\varepsilon - \varepsilon_a)}{\rho(\varepsilon)}}\chi_a. \quad (10)$$

The four orbital are then rotated to the original atomic basis, $i_\varepsilon = S^{-\frac{1}{2}}U\bar{\iota}(\varepsilon)$ and $a_\varepsilon = S^{-\frac{1}{2}}U\bar{a}(\varepsilon)$. Finally, the two-electron integrals $\langle ij|ab\rangle$ involving the four random orbitals are performed on a 3D real-space grid, applying fast Fourier techniques for the convolution operation.

There are several sources for numerical errors in estimation of the MP2 energy in a given basis. The systematic errors result from the parameter $\gamma$ in the squared-Lorentzian but this can be controlled to a desired predefined accuracy. Stochastic errors (SE) are due to random fluctuations and can be controlled by repeated sampling. After $I$ samplings (iterations) the SE is equal to $SE_1/\sqrt{I}$ where $SE_1$ is the "SE per iteration".

We now describe applications of the method for small and medium sized molecules, and for huge fullerene clusters (with 36 and 64 fullerenes) with over 15,000 basis functions. We used the Siesta package [20] within the SZ, DZ and DZVP Siesta basis sets. The results of these MP2 energy calculations are summarized in Table 1. For small molecules, we compare the results of the stochastic calculations with the explicit summation results of Eq. (1), referred to as 'deterministic'. We provide details about the basis set used, the total number of basis function, the number of stochastic sets of orbitals used (iterations), the statistical error in the MP2 energy per electron (SE), the SE per iteration and whether the guiding weight function was used or not.

We find that the MP2 correlation energy per electron is roughly independent of the system size for systems with similar electronic character. This is certainly the case for Benzene, Naphthalene and Pentacene series ($E_{MP2}/N \approx 0.52 - 0.56$ eV) and also for varying sizes of Fullerene clusters ($E_{MP2}/N \approx 0.68 - 0.70$ eV).

The SE per iteration decreases somewhat with the size of the system. For example, the SE per iteration decreases from 2.7 eV to 2.0 eV when a fullerene is replaced by a fullerene cluster. Similarly, the SE per iteration reduces from 5.6 eV to 3.7 eV going from Benzene to Pentacene. This is in contrast to the time-domain algorithm, where the SE decreased and then leveled off for a linear chain model and increased and leveled off for a 3D silicon nanocrystal (see more below). The reduction of the SE per iteration indicates that the energy formulation of MP2 benefits from self-averaging, again in contrast to the time-domain approach. Furthermore, as the basis increases the SE per iteration increases from 2.7eV to 7.1eV, which is natural since the MP2 energy also increases by a similar factor.

Table 1: MP2 energies (all energies are per electron, in eV) using Siesta basis sets with the Stochastic Energy-domain approach for a range of molecules, up to clusters of 36 and 64 fullerenes (with center-to-center distances of 10 Angstrom). For small molecules we compare the stochastic results with the traditional explicit summation results. The SE per iteration is the SE multiplied by the square root of the number of iterations. Catechol-Fullerene refers to a fullerene derivative with a catechol molecule fused to an open fullerene cage through a 2-carbon linker.

| Molecule | $N_e$ | Basis set | $M$ | $I$ | $E_{MP2}/N_e$ | | SE | $\sqrt{I} \times SE$ | Guiding Function? |
|---|---|---|---|---|---|---|---|---|---|
| | | | | | Deterministic | Stochastic | | | |
| Water | 8 | SZ | 6 | 1048576 | 0.197 | 0.196 | 0.002 | 1.5 | N |
| | 8 | DZVP | 23 | 1048576 | 1.015 | 1.029 | 0.012 | 12.6 | N |
| Ethylene | 12 | SZ | 12 | 4194304 | 0.447 | 0.454 | 0.006 | 12.7 | N |
| Benzene | 30 | SZ | 30 | 4194304 | 0.521 | 0.523 | 0.005 | 11.1 | N |
| | 30 | SZ | 30 | 1048576 | 0.521 | 0.522 | 0.005 | 5.6 | Y |
| Naphthalene | 48 | SZ | 48 | 1048576 | 0.558 | 0.557 | 0.004 | 4.5 | Y |
| Pentacene | 102 | SZ | 102 | 1048576 | | 0.560 | 0.004 | 3.7 | Y |
| Catechol-Fullerene | 290 | SZ | 286 | 524288 | | 0.709 | 0.004 | 2.7 | Y |
| | 290 | DZ | 572 | 524288 | | 0.995 | 0.005 | 3.6 | Y |
| | 290 | DZVP | 940 | 524288 | | 1.442 | 0.009 | 6.8 | Y |
| Fullerene | 240 | SZ | 240 | 1048576 | | 0.678 | 0.003 | 2.7 | Y |
| | 240 | DZ | 480 | 1048576 | | 0.981 | 0.005 | 5.1 | Y |
| | 240 | DZVP | 780 | 1048576 | | 1.434 | 0.007 | 7.1 | Y |
| Fullerene $3 \times 3 \times 4$ | 8640 | SZ | 8640 | 131072 | | 0.704 | 0.006 | 2.2 | Y |
| Fullerene $4 \times 4 \times 4$ | 15360 | SZ | 15360 | 65536 | | 0.696 | 0.008 | 2.0 | Y |

In Table 2 we show in more detail, the effect of the guiding function on the SE for a set of molecules. It reduces the SE per iteration by approximately 2 for the smallest molecule and nearly a factor of 4 for the largest, implying reduction of the number of stochastic orbitals required to achieve a given SE by a factor 4 for smaller molecules and 16 for larger ones.

Table 2: The effects of the guiding function. The guiding function reduces the dispersion in the MP2 energies by a factor of 2-4, thereby reducing the number of iterations by 4-16. For all cases $I = 10^6$.

| Molecule | Basis Set | Guiding function? | $\frac{E_{MP2}}{N_e}$ | SE | $\sqrt{I} \times SE$ |
|---|---|---|---|---|---|
| Benzene | SZ | N | -0.518 | 0.010 | 10.1 |
| | | Y | -0.522 | 0.005 | 5.6 |
| Naphthalene | SZ | N | -0.548 | 0.009 | 8.8 |
| | | Y | -0.557 | 0.004 | 4.5 |
| Pentacene | SZ | N | -0.583 | 0.015 | 15.8 |
| | | Y | -0.560 | 0.004 | 3.7 |
| Fullerene | SZ | N | -0.700 | 0.008 | 8.6 |
| | | Y | -0.678 | 0.003 | 2.7 |
| | DZ | N | -0.972 | 0.013 | 13.7 |
| | | Y | -0.980 | 0.005 | 5.1 |
| | DZVP | N | -1.448 | 0.023 | 23.6 |
| | | Y | -1.433 | 0.007 | 7.1 |

The principles of our energy-domain stochastic approach can also be applied to a real-space-grid or plane-waves representation. Here, the underlying basis is orthogonal so there is no overlap matrix to consider, on the other hand the Hamiltonian matrix is too large to be diagonalized and so iterative sparse matrix techniques must be applied. The random orbitals (cf. Eqs. (3) and (4)) can be obtained similarly to in ref. [13] by expanding $\sqrt{\delta_\sigma(\varepsilon - H)}$ as a Chebyshev expansion.[21] From the structure of the Chebyshev series it is possible to obtain several stochastic orbitals of different $\varepsilon$ from a single expansion and we use this property to obtain 4 unoccupied and 16 occupied orbitals. In addition, the density of states $\rho(\varepsilon)$ is calculated separately using a stochastic trace formula as in reference.[19] Finally, the two electron integrals for the orbitals on the grid are obtained using fast Fourier convolution techniques.

We apply the energy-domain stochastic method to hydrogen passivated spherical silicon nanocrystals (NCs) of several sizes. We use a semi-empirical pseudopotential model to construct the single particle Hamiltonian[22] and a real-space grid to represent the single particle orbitals.[23] In Table 3 we summarize the results for three systems sizes: $Si_{35}H_{36}$, $Si_{87}H_{76}$, and $Si_{353}H_{196}$. The total number of electrons varies from 176 to 1608 and the size of the Hamiltonian matrix from $32^3$ to $64^3$. Since a direct calculation of the MP2 correlation energy is prohibited for these NCs, we compare the current approach to our previous time-domain stochastic approach.[13]

The agreement between the energy- and time-domain stochastic approaches is excellent for the smaller NCs (differences are well within the SE). For the largest NC, we find small deviations between the two approaches, which may result from systematic errors introduced by the finite Chebyshev expansion length. For the smaller NC, we find that within the SE, 2048 Chebyshev terms are sufficient to converge the MP2 correlation energy to within 0.04 eV/atom. For the larger NC, however, one would require a longer Chebyshev expansion since the quasi-particle gap is smaller by nearly a factor of 2 compared to the smaller NC. Tests of the effect of the length of the Chebyshev series for the larger NC require more work and larger computational resources.

Table 3: Comparison between the energy (present) and time (ref. [13]) domain stochastic MP2 calculations for hydrogen passivated silicon nanocrystals. The number of stochastic orbitals used is 50,000 for the energy-domain results and 2,500 for the time-domain results.

| | Nanocrystal: | $Si_{35}H_{36}$ | $Si_{87}H_{76}$ | $Si_{353}H_{196}$ |
|---|---|---|---|---|
| | $N_e$ | 176 | 424 | 1608 |
| | $M$ | 32768 | 110592 | 262144 |
| $\frac{E_{MP2}}{N_e}$ | Energy-domain | -1.03 | -1.07 | -1.23 |
| | Time domain | -1.06 | -1.13 | -1.31 |
| SE | Energy-domain | 0.03 | 0.04 | 0.04 |
| | Time domain | 0.02 | 0.03 | 0.04 |
| $\sqrt{I} \times$ SE | Energy-domain | 6.8 | 8.7 | 8.6 |
| | Time domain | 0.8 | 1.3 | 1.8 |
| Ratio of computational work | | 4.8 | 3.3 | 1.4 |

It is interesting to note that the SE per iteration in the energy-domain calculations, which does not incorporate the guiding function, are much larger in comparison to the time-domain approach. However, since the time-domain approach requires in addition to the filtering step, a propagation step, the overall ratio of computational work is between 1.4 and 4.8 depending on the size of the NC. For the larger NC, this ratio should be multiplied by 2 if an appropriate length of the Chebyshev expansion is used. On the other hand we did not use a guiding function in these calculations, and that would have allowed us to reduce the energy-domain computational effort by a factor of 2-4 based on the results reported in Table 2.

Table 4: The MP2 correlation energy and SE for $Si_{35}H_{36}$ for different values of the Chebyshev expansion length.

| Expansion length: | 2048 | 4096 | 8192 |
|---|---|---|---|
| Stochastic MP2 energy | -1.066 | -1.029 | -1.085 |
| SE | 0.04 | 0.03 | 0.03 |
| SE per iteration | 9.2 | 6.8 | 6.8 |

To summarize, we developed an energy-domain stochastic method for estimating the MP2 energy which gives converged per-electron properties. For a basis of contracted Gaussian functions (CGF), the energy-domain approach is more suitable than our previous time domain approach. A key element is the introduction of a guiding function which we find to reduce the computational effort by a factor of 4-16. The approach is also suitable for a real-space-grid or plane-waves representation where the time domain approach seems more suitable, but the energy-domain approach seems to work nearly as fast when the system size increases.

Our results show that it is feasible to perform MP2 correlation energy calculations even for very large systems. For CGF basis sets, assuming that the Hartree-Fock orbitals and orbital energies are available, our MP2 approach scales as $\mathcal{O}(N \log N)$. In the real-space-grid or plane-waves application the unoccupied orbitals and energies are not known and we rely on application of filters to random wave functions (applied using Chebyshev expansions) and the scaling is also $\mathcal{O}(N \log N)$.

QG is grateful for support from the Fellowship for Students of Basic Subjects (Zhejiang University) during a stay in UCLA. YG and DN were supported by NSF grant CHE-1112500. RB and DN were supported by the US-Israel Binational Foundation (BSF). ER would like to thank the Israel Science Foundation (grant number 611/11) for financial support and the Marko and Lucie Chaoul Chair. We thank Chris Arntsen for valuable discussions and Rob Thompson for providing the structure for the Catechol-Fullerene derivative.